\documentclass[sigconf]{acmart}

\AtBeginDocument{%
  \providecommand\BibTeX{{%
    \normalfont B\kern-0.5em{\scshape i\kern-0.25em b}\kern-0.8em\TeX}}}

\setcopyright{acmcopyright}
\copyrightyear{2018}
\acmYear{2018}
\acmDOI{XXXXXXX.XXXXXXX}

\acmConference[Conference acronym 'XX]{Make sure to enter the correct
  conference title from your rights confirmation email}{June 03--05,
  2018}{Woodstock, NY}
\acmPrice{15.00}
\acmISBN{978-1-4503-XXXX-X/18/06}

\usepackage{amsmath}
\usepackage{algorithm}
\usepackage{algpseudocode}
\usepackage{multirow}
\usepackage{float}
\usepackage{enumitem}

\begin{document}

\title{Better Late Than Never: Formulating and Benchmarking Recommendation Editing
}

\author{Chengyu Lai}
\email{laichengyu@zju.edu.cn}
\affiliation{%
  \institution{Zhejiang University}
  \city{HangZhou}
  \country{China}
}

\author{Sheng Zhou}
\email{zhousheng\_zju@zju.edu.cn}
\affiliation{%
  \institution{Zhejiang University}
  \city{HangZhou}
  \country{China}
}

\author{Zhimeng Jiang}
\email{zhimengj@tamu.edu}
\affiliation{%
  \institution{Texas A\&M University}
  \city{College Station}
  \country{USA}
}

\author{Qiaoyu Tan}
\email{qiaoyu.tan@nyu.edu}
\affiliation{%
  \institution{New York University Shanghai}
  \city{Shang Hai}
  \country{China}
}

\author{Yuanchen Bei}
\email{yuanchenbei@zju.edu.cn}
\affiliation{%
  \institution{Zhejiang University}
  \city{HangZhou}
  \country{China}
}

\author{Jiawei Chen}
\email{sleepyhunt@zju.edu.cn}
\affiliation{%
  \institution{Zhejiang University}
  \city{HangZhou}
  \country{China}
}

\author{Ningyu Zhang}
\email{zhangningyu@zju.edu.cn}
\affiliation{%
  \institution{Zhejiang University}
  \city{HangZhou}
  \country{China}
}

\author{Jiajun Bu}
\email{bjj@zju.edu.cn}
\affiliation{%
  \institution{Zhejiang University}
  \city{HangZhou}
  \country{China}
}







\renewcommand{\shortauthors}{Trovato and Tobin, et al.}

\begin{abstract}
Recommendation systems play a pivotal role in suggesting items to users based on their preferences. However, in online platforms, these systems inevitably offer \textit{unsuitable} recommendations due to limited model capacity, poor data quality, or evolving user interests. Enhancing user experience necessitates efficiently \textit{rectify} such unsuitable recommendation behaviors. This paper introduces a novel and significant task termed \textit{recommendation editing}, which focuses on modifying \textit{known and unsuitable} recommendation behaviors. Specifically, this task aims to adjust the recommendation model to eliminate known unsuitable items without accessing training data or retraining the model. We formally define the problem of recommendation editing with three primary objectives: \textit{strict rectification, collaborative rectification}, and \textit{concentrated rectification}. Three evaluation metrics are developed to quantitatively assess the achievement of each objective. We present a straightforward yet effective benchmark for recommendation editing using novel Editing Bayesian Personalized Ranking Loss. To demonstrate the effectiveness of the proposed method, we establish a comprehensive benchmark that incorporates various methods from related fields. Codebase is available at \textcolor{blue}{\url{https://github.com/cycl2018/Recommendation-Editing}}.


\end{abstract}

\ccsdesc[500]{Information systems~Recommender systems}

\keywords{Recommender Systems, Collaborative Filtering, Model Editing}



\maketitle

\section{Introduction}
Recommender systems (RSs) provide personalized recommendations by mining user preferences for items and have
achieved great success in various applications~\cite{gao2023survey,chen2023bias}.
Existing recommendation systems typically mine potential preferences from massive user historical behavior data, such as clicks and purchases,  and then serve in online systems to recommend items. However, due to insufficient/unclean user behavior data~\cite{wang2021denoising}, limited model capabilities~\cite{fan2023recommender} as well as evolving user interests~\cite{zhang2020retrain}, recommendation systems may provide unsuitable recommendations to users and also reported by users.
Among these unsuitable recommendation items, some may be uninteresting to users, while others can be more severe, causing user aversion or violating relevant laws and regulations. 
For example, recommending adult products to anonymous underage users or suggesting erroneous information to individuals with different ethnic or religious beliefs. Addressing this entails several requirements: (1) rapid rectification to mitigate negative impacts; (2) computation-efficient rectification to handle the potential frequent errors; (3) Preferably, solutions should work well even without training data access—a major plus for privacy—though access is allowed.
Therefore, developing methods to quickly and efficiently rectify erroneous recommendations is crucial for fostering responsible and sustainable recommender systems in the real world.

\begin{figure}[t]
     \centering
     \includegraphics[width=0.92\linewidth, trim=0cm 0cm 0cm 0cm,clip]{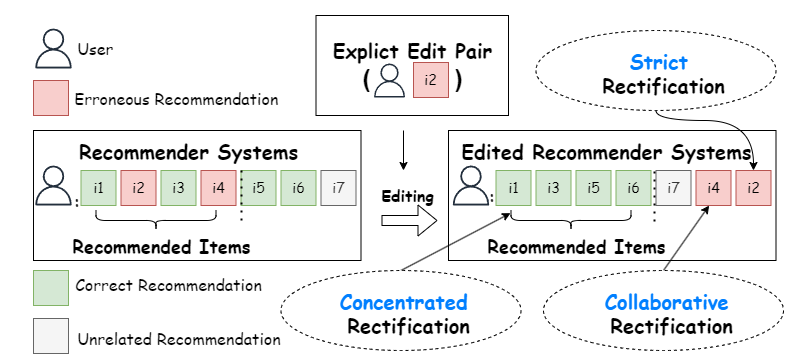}
     \caption{Illustration of Recommendation Editing.}
     \label{fig:editing}
     \vspace{-7mm}
\end{figure}

To achieve recommendation rectification, we introduce \textbf{recommendation editing} task with a systematic framework, as shown in Figure~\ref{fig:editing}. Given an explicit edit user-item pair, recommendation editing can edit RS model with desired recommendation results. To fully understand and implement recommendation editing, it is essential to define three key goals:
(1) \textbf{Strict Rectification}: unsuitable recommendation items, particularly those leading to serious issues like discrimination or illegal content, imperatively be removed after editing. 
Traditional methods like online/incremental recommendation~\cite{zhang2023survey} or approaches considering negative feedback~\cite{huang2023negative} aim to iteratively refine and enhance the model with ongoing new data rather than strictly rectify known and unsuitable recommendation behavior without model retraining.
Despite their infrequency, the significant negative impact of erroneous recommendations necessitates their prompt rectification.
(2) \textbf{Collaborative Rectification}: the similar but unobserved erroneous recommendations items also imperatively be removed after editing. Given the significant negative impact of such recommendations, swift recommendation adjustments are necessary to correct these errors comprehensively and prevent similar future errors. While efficient fine-tuning strategies~\cite{wang2023comprehensive} can address specific observed errors, they risk overfitting and may not generalize well to rectifying unobserved but related errors.
(3) \textbf{Concentrated Rectification}: the majority of appropriate recommendations are preserved after editing, ensuring that unsuitable items from recommendations imperatively be removed only when necessary.  Although addressing erroneous recommendations is important, large-scale, mature recommendation systems usually contain a small fraction of these severe errors. Retraining the model with recommendation error correction constraints can meet the above needs but might compromise the quality of the vast majority of acceptable recommendations. This compromise could detract from the overall user experience and jeopardize the system's long-term viability.

Although model editing is widely adopted in natural language processing (NLP) and computer vision (CV)~\cite{yao2023editing,wang2023knowledge} to rectify model behaviors, it is a complex task in RS to rectify erroneous recommendations, and these solutions are not directly applicable to recommendation systems due to two key considerations. Firstly, the inherent nature of recommendation systems is non-iid(Non-Independent and Identically Distributed) data and involves a dynamic interaction between users and items, necessitating a sophisticated approach to loss functions that accurately captures this relationship. Secondly, the core objective of recommendation systems is to prioritize and rank items effectively, which requires the consideration of specialized ranking metrics. This distinction highlights the unique challenges in enhancing recommendation systems to avoid unsuitable suggestions, underlining the importance of a tailored approach for `recommendation editing' that addresses these specific needs without the necessity for model retraining.




To bridge the gap, this study concentrates on improving and evaluating the performance of recommendation editing. Our investigation unfolds across three critical dimensions. Firstly, we delve into the recommendation editing problem formulation and \textit{three} evaluation metrics designed to gauge the effectiveness of recommendation editing from various perspectives in Section~\ref{sect:evaluation}. Subsequently, in Section~\ref{sect:method}, we explore the editing loss functions, a simple yet effective strong baseline, tailored for recommendation editing. 
Finally, a comprehensive benchmark with various baselines borrowing from other domains is thoroughly examined in Section~\ref{sect:exp}. We demonstrate the effectiveness of our proposed methods after editing, thereby mitigating the issue of unsuitable recommendations. Our work aims to bridge the gap in current research by offering a comprehensive benchmark of RSs' performance when tasked with the complex challenge of recommendation editing. We outline the major contributions below. 
\begin{itemize}[topsep=0pt,partopsep=0pt, leftmargin=*]
\item This study introduces the concept of recommendation editing to recommendation systems, aiming to rectify erroneous recommendations made by well-trained models. It highlights the significance of addressing this issue for the development of responsible recommendation systems.
\item We have developed three evaluation metrics specifically designed to measure the performance of recommendation editing. These metrics enable a comprehensive assessment of its effectiveness.
\item Our research benchmarks a series of recommendation editing methods, most of which draw inspiration from model editing techniques in the NLP field. Additionally, we propose a simple yet effective baseline method, named EFT, for recommendation editing and conduct a comparative analysis of various approaches.
\end{itemize}

\section{Related Works}

\subsection{Online Recommendation}
With the exponential growth of online information, recommender systems have become ubiquitous in various applications, helping users discover items aligned with their interests~\cite{koren2021advances,he2017neural,wang2019neural,he2020lightgcn}. In practice, RS operates in an incremental manner as new users, items, and user-item interactions are constantly being observed over time.

\textit{Online recommendation.} To meet this demand, in recent years, some works have started focusing on efficient model updates for online learning in recommender systems.
Among them, one kind of approach is based on continual learning, which typically requires maintaining a dynamic user model and making real-time adjustments based on user feedback and new data. ADER~\cite{mi2020ader} periodically replays previous training samples to the current model with an adaptive distillation loss.
Further, ReLoop~\cite{cai2022reloop} builds a self-correction learning loop for recommender systems with customized optimization loss based on user feedback.
Another kind of approach is based on incremental learning, of which the goal is to quickly adapt current models to new data, and maintain high performance with minimal computational overhead through effective incremental learning algorithms or update strategies.
MAN~\cite{mi2021memory} enhances a neural recommender system by incorporating a continuously queried and updated nonparametric memory.
Then, FIRE~\cite{xia2022fire} proposes a fast incremental learning method from the graph signal processing perspective.

\textit{Incremental recommendation.} Existing online learning methods for recommendation rely on \textit{explicit user feedback}. However, when containing incorrect items, it is preferable to proactively edit recommendations and ensure that incorrect items are removed before being recommended to the users, rather than expecting users to provide negative feedback after being exposed to such items.



\subsection{Recommendation with Negative Feedback}
The vast majority of recommendation systems make suggestions based on users' historical positive feedback, neglecting users' negative feedback. However, negative feedback is widely prevalent in real-world scenarios, such as low ratings, content skipping, and clicking the dislike button~\cite{huang2023negative}. It has been pointed out that utilizing negative feedback information can significantly enhance the performance of recommendation systems~\cite{jeunen2019revisiting}. To fully leverage the rich negative feedback, some methods have begun to explore how to combine positive and negative feedback for recommendations. For instance, movie
recommendation systems~\cite{chen2021movie} are proposed to establish a user positive profile and negative profile to enhance movie recommendations. Meanwhile, SiReN~\cite{seo2022siren} and SiGRec~\cite{huang2023negative} utilize both positive and negative feedback to construct a Signed Graph. By employing graph-based methods to integrate positive and negative feedback, these approaches aim to improve the performance of recommendation systems.

\subsection{Model Editing}
The concept of model editing was first introduced in the context of generative adversarial networks~\cite{bau2020rewriting}. 
Model edits were achieved by treating a neural network layer as a linear associative memory, allowing for the derivation of a low-rank update to the layer weights. This update was based on an input or a set of inputs, commonly referred to as a  ``concept''.

In addition, a variety of alternative editing algorithms have been proposed. For instance, \cite{de2021editing,mitchell2022fast} trained hyper-networks to update model weights.
\cite{dai2022knowledge} identified ``knowledge neurons" and then surgically modified them.
\cite{mitchell2022memory} augmented a target model with an auxiliary detector trained to identify inputs relevant to the editing task, and then diverted them to an edited model.
See also \cite{hase2021language} for an analysis of beliefs stored in language models and methods to update those beliefs. 
Recent work also extended model editing to open vocabulary image-language models (e.g., CLIP~\cite{radford2021learning}), where it was demonstrated that interpolating between the original and edited weights allows users to navigate a trade-off between performance on the original and edited tasks~\cite{ilharco2022patching}.

The particular relevance to our work is the recent findings in \cite{lee2023surgical}: updating only a subset of model weights can result in greater robustness to distribution shift. We consider a different notion of robustness in this work, as well as a different set of models and editing/fine-tuning tasks than that work. We likewise update a more granular subset of weights (\cite{lee2023surgical} updates ``blocks" of weights, whereas we typically update weights in a single layer).

\section{Recommendation Editing}

In this section, we will begin by introducing the key notations used in this paper.
Subsequently, we will formulate the recommendation editing problem and propose comprehensive evaluation metrics that consider the three properties of editing.
Lastly, we will delve into a detailed discussion of the unique challenges posed by the recommendation editing problem.


\subsection{Problem Definition}
\label{sec-pd}

\textbf{Vanilla Recommendation.}
In vanilla recommendation, we are given a user set $\mathcal{U}$ and an item set $\mathcal{I}$.
Let $u$ (or $i$) denote a user (or an item) in $\mathcal{U}$ (or $\mathcal{I}$). 
Let $\mathcal{M}:=\mathcal{U} \times \mathcal{I}\rightarrow \{0,1\}$ be the recommendation model, which predicts whether a user actually likes an item or not.
The recommender system takes the historical feedback data for training, which can be formulated as a set of interaction pairs $\mathcal{D}_{train}=\{(u,i,y)_{m}\}$ where $(u,i,y)_{m}$ denotes the $m$-th labeled pair.
The recommender system aims at capturing user preference and learning the label function $f$ from the training data $\mathcal{D}_{train}$ so that the item $i$ that user $u$ is interested in can be recommended $y_{ui}=1$ while the item $k$ that not interested in is not recommended $y_{uk}=0$. 
After the recommender system is trained, it is deployed and makes predictions for users in the test data $\mathcal{D}_{test}$.

\textbf{Recommendation Editing.}
Given the trained recommender system $f$ and the test data $\mathcal{D}_{test}$, the recommendation results can be denoted as $\mathcal{R}=\{(u,i,\hat{y})_{r}\}$, where $(u,i,\hat{y})_{r}$ denotes the $r$-th recommendation results.
However, some of the recommendation results may be inappropriate for certain users (such as different ages, races, or disabilities), which deserve to be edited to prevent negative impacts on users.
We denote $\mathcal{E}=\{(u,i,\tilde{y})_{e}\}\subset \mathcal{R}$ as the set of user-item pairs that are expected to be edited, where $(u,i,\tilde{y})_{e}$ denotes the $e$-th editing pair. 
For same user/item pair $(u,i)_{r}=(u,i)_{e}$, the label is different $\hat{y}_{r}\neq \tilde{y}_{e}$. 

To understand recommendation editing setting, we highlight the following three types of pairs: (1) \textbf{Explicit Editing Pairs $\mathcal{E}^{\mathbb{E}}$} refer to the few accessed editing pairs that are expected to be strictly edited. In practice, they can be collected by internal testing, users' complaints, and feedback from regulatory authorities. (2) \textbf{Implicit Editing Pairs $\mathcal{E}^{\mathbb{I}}$} refer to the pairs that are similar to the explicit editing pairs $\mathcal{E}^{\mathbb{E}}$ and should be edited. However, they are not accessed in the editing process. In practice, they can be split from the ground truth Editing pairs $\mathcal{E}$ or inferred from the few explicit editing pairs $\mathcal{E}^{\mathbb{E}}$. We will introduce the detailed splitting of three sets of pairs in the experimental part. (3) \textbf{Unnecessary Editing Pair $\bar{\mathcal{E}}$} refer to the pairs that are different from the editing pairs $\mathcal{E}$ and expect to be not edited so that the most ordinary users' experiences will not be affected. To summarize, the above set of pairs satisfies:
\begin{equation}
    \mathcal{E}^{\mathbb{E}} + \mathcal{E}^{\mathbb{I}} = \mathcal{E} \quad \text{and} \quad \mathcal{E}^{\mathbb{E}} + \mathcal{E}^{\mathbb{I}} + \bar{\mathcal{E}} = \mathcal{R}.
\end{equation}
Due to privacy protection and sparsity of user behavior, we often do not have access to the complete set of editing samples, instead, we only have access to a small explicit subset $\mathcal{E}^{\mathbb{E}}\subset \mathcal{E}$.
The recommendation editing aims to \textbf{editing the recommendation output without retraining the entire model} so that the following three properties can be satisfied:
\begin{itemize}[leftmargin=*]
    \item \textbf{Accurate Editing}: The primary task of recommendation editing is to accurately edit the inappropriate recommendation results $(u, i, \hat{y})_e\in\mathcal{E}^{\mathbb{E}}$ can be edited rigorously, thereby avoiding further negative impacts.
    \item \textbf{Collaborative Editing}: Due to the difficulty in obtaining all the samples that require editing (otherwise, we could directly use rules to modify the recommendation output), recommended editing aims to collaboratively edit pairs of similar users and items $(u, i, \hat{y})_i\in\mathcal{E}^{\mathbb{I}}$ as well, in order to reduce potential negative impacts.
    \item \textbf{Prudent Editing}: While recommendation editing is effective, the samples that need to be edited (explicitly or implicitly) typically represent only a small fraction of all the sample pairs. Otherwise, it would be preferable to retrain the model. Therefore, model editing should be prudent and preserve recommendation results for $(u, i, \hat{y})_i\in\bar{\mathcal{E}}$.
\end{itemize}

Note that in practice, the recommender system is usually trained on large-scale user behavior data and has a large number of parameters, we can not re-train or fine-tune the model since we do not have access to the model parameters as well as training data.



\subsection{Editing Evaluation Metrics}\label{sect:evaluation}
Based on the above properties, we propose a comprehensive evaluation approach to assess the effectiveness of the recommendation editing. Specifically, we propose three metrics to evaluate the three aforementioned proprieties of recommendation editing:

\begin{itemize}[leftmargin=*]
    \item \textbf{Editing Accuracy (EA)} measures the ratio of explicit editing pairs been successfully edited, which can be formulated as:
    \begin{equation}
        EA = \frac{|\{r^{\mathbb{E}}_{ui}> k\}|}{|\mathcal{E}^{\mathbb{E}}|}, \quad (u,i)\in \mathcal{E}^{\mathbb{E}}.
    \end{equation}
    where $k$ is the number of recommended items, $r^{\mathbb{E}}_{ui}$ is the rank of item $i$ among user $u$'s recommendations after editing, $|\cdot|$ denotes the number of samples in a set. Note that the editing accuracy only measures the success rate of editing in explicit editing pairs $\mathcal{E}^{\mathbb{E}}$. Given these explicit pairs, the fundamental requirement of recommendation editing is to ensure all of them are edited.
    \item \textbf{Editing Collaboration (EC)} measures the ratio of implicit editing pairs been successfully edited, which can be formulated as:
\begin{equation}
\begin{aligned}
    EC=& \frac{1}{|\mathcal{E}^{\mathbb{I}}|}|\{r^{\mathbb{E}}_{ui}>k\ \& \ r_{ui}\leq k\}|  \\  & + |\{r^{\mathbb{E}}_{ui}>r_{ui}\ 
\& \ r_{ui}> k\}|, \quad (u,i)\in \mathcal{E}^{\mathbb{I}}.
\end{aligned}
\end{equation}

    where $r_{ui}$ is the rank of item $i$ among the recommendations for user $u$ before editing. 
    It is important to highlight that the pairs in the implicit editing pairs are divided into two groups based on their ranks before editing: 
    For pairs that were initially in the top-$k$ recommendations, the editing is considered successful only if the pair is no longer in the top-$k$ recommendations after editing.
    For pairs that were not in the top-$k$ recommendations before editing, the editing is considered successful if the rank of the pair decreases after editing. We leave the possibility of a more detailed subdivision for future research.
    \item \textbf{Editing Prudence (EP)} measures the ratio of unnecessary editing pairs not been edited, which can be formulated as:
    \begin{equation}
        EP = \frac{|\{r_{ui}< k\} \cap \{r^{\mathbb{E}}_{ui}< k\} |}{|\{r_{ui}< k\} \cup
\{r^{\mathbb{E}}_{ui}< k \}| }, \quad (u,i)\in \bar{\mathcal{E}}.
    \end{equation}
where $\cap$ and $\cup$ are intersection and union operations on sets. In other words, recommendation systems should consistently avoid suggesting unrelated items to users.
\end{itemize}
To measure the overall erroneous editing performance, note that the number of implicit and unnecessary editing pairs ($\mathcal{E}^{\mathbb{I}}$ and $\bar{\mathcal{E}}$) could be imbalanced,  we propose \textbf{Editing Score (ES)}, the harmonic mean between EC and EP, to evaluate erroneous editing performance, i.e. $ES=2\frac{EC\cdot EP}{EC + EP}$ \footnote{The harmonic mean in ES is similar to F1 penalizing large disparities between precision and recall, ensuring that both metrics need to be high to achieve a high score.}.




\subsection{Discussions}
Given the definition and the three key goals of recommendation editing, in this subsection, we discuss the relations and differences between recommendation editing with many existing methods.

\textit{Online/Incremental Recommendation} aims to continuously update the recommendations model based on the new users' feedback. While online/incremental recommendation methods can be adapted to handling negative feedback, the goals are quite different from model editing. Specifically, online/incremental recommendation focuses on model adaption to new user feedback for accurate recommendations for all users rather than editing those important but erroneous recommendation items. Consequently, the erroneous recommendation items that require editing may be compromised using online/incremental recommendation methods due to the preference modeling of a massive number of users. 
Model Editing sets strictly rectifying erroneous recommendation items as the primary goal, while online aims at adapting the new user feedback.

\textit{Recommendation with Negative Feedback} focuses on modeling user preferences by considering both positive and negative feedback. However, its effectiveness relies on training from scratch and thus requires access to all historical feedback data. This requirement is impractical in recommendation editing contexts, due to the inherent inefficiencies and the common issue of incomplete datasets. 

\textit{Recommendation unlearning} seeks to eliminate the influence of specific training data from the model, whereas recommendation editing aims to avert erroneous recommendations generated by the recommendation model. 
The divergent goals of these approaches pose challenges in applying recommendation unlearning strategies to meet the strict rectification goal of recommendation editing.


\textit{Model Editing} has garnered attention in computer vision and natural language processing.
However, it overlooks the inherent nature of recommendation systems involving non-independent and identically distributed (non-iid) data and dynamic interactions between users and items. Furthermore, model editing primarily targets instance-wise tasks, such as classification. In recommendation editing, the ranking task involves the relative importance of items for each user. This distinction emphasizes the need for recommendation editing approaches specifically designed to address the nuanced challenges of ranking.

\section{Recommendation Editing Loss}\label{sect:method}


As we have discussed earlier, existing works can not fully satisfy the recommendation editing requirements.
To break this impasse, in this section, we aim to propose a simple yet effective recommendation editing loss that can be applied to most existing recommendation models.

Among existing recommendation methods (e.g. matrix factorization or deep learning methods), most of them use different techniques to learn user embedding $\mathbf{P}$ and item embedding $\mathbf{Q}$.
The user-item score is usually calculated by the inner product between user embedding and item embedding:
\begin{align}
    x_{u,i}=P_u^TQ_i,
\end{align}
where $x_{u,i}$ is the score between user $u$ and item $i$.
The final recommendation set $\mathcal{R}_{u}$ for user $u$ is selected as the items with top-$k$ highest scores among all the items.
To achieve the goal of model-agnostic recommendation editing, we propose to directly operate on user embedding $\mathbf{P}$ and item embedding $\mathbf{Q}$, regardless of the way that learn these embeddings. It is worth noting that this loss function can also be directly used to optimize the whole recommendation model by backpropagating the gradients to the model parameters that obtain the representations.

In existing recommendation methods, Binary Cross-Entropy (BCE) and Bayesian Personalized Ranking (BPR) are the two most commonly used loss functions.
The BCE loss is typically designed for the binary classification task. Traditional recommendation methods treat the user-item pair as a binary classification task where $y_{ui}=1$ denotes item $i$ is interesting for user $u$, and $y_{ui}=0$ otherwise. In the recommendation editing scenario, the editing pair provides the label $y_{ui}=0$ for the user-item pair. It is natural to optimize the corresponding user embedding $P_{u}$ and item embedding $Q_{i}$ by pushing them apart. 
However, in this way, we are only able to optimize and update the user embedding $P_{u}$ and item embedding $Q_{i}$, while the embedding of all the other users and items is not updated.
As a result, only the item ranking for user $u$ and ranking of item $i$ in all other users will be updated, and the other user-item pairs: $\{(u', i'): u' \neq u \ \text{and} \ i' \neq i\}$ will not be updated.
Concerning the comprehensive requirements of recommendation editing, although using the BCE loss may successfully edit the explicit editing pairs (high editing accuracy), the failure in editing implicit editing pairs $\mathcal{E}^{\mathbb{I}}$ makes it not suitable for recommendation editing.

The BPR loss~\cite{rendle2012bpr} is typically designed to optimize the relative ranking relationships between items for each user based on their preferences. 
In recommendation systems, we are more interested in the ranking of items by users rather than just binary classification predictions.
As a result, we tend to utilize the idea of BPR loss to help edit the recommender system.
In traditional recommendation, the explicit user feedback $(u,i)$ (such as user $u$ clicks or consumes an item $i$) is treated as a positive item for the user $u$, while the negative pairs $(u,j)$ are usually sampled from implicit user feedback.
The recommendation methods are usually optimized by maximizing the likelihood of correctly assigning a higher score $x_{u,i}$ for the positive pair than other negative pairs $x_{u,j}$.
In recommendation editing, given a few editing pairs $(u,i)$, it is natural to treat them as negative pairs.
The core of utilizing BPR loss for editing lies in the definition of positive samples.



When considering any $(u_e, i_e) \in \mathcal{E}^\mathbb{E}$, we generate recommendation results for $u_e$ using the k items with the highest scores:
\begin{align}
    \mathcal{R}_{u_e}=\{i | x_{u_e , i} \in TopK(X_{u_e})\},
\end{align}
where $X_{u_e}$ is the set of scores for all items by $u_e$.

Since our goal is to ensure that $i_e$ does not appear in $\mathcal{R}_{u_e}$, while also minimizing changes to the model's original recommendations for other items, this suggests that we need to lower the rating of $u_e$ for $i_e$ while maintaining high ratings for other items. Therefore, we achieve the goal of model editing by reinforcing the ranking of other items in $\mathcal{R}_{u_e}$ above $i_e$. 
In other words, we treat $i_e$ as a negative sample for the BPR loss, with the other items in $\mathcal{R}_{u_e}$ considered as positive samples. Our proposed E-BPR (Editing BPR) loss can be formalized as:
\begin{align}
    \mathcal{L} =\sum_{(u_e,i_e) \in \mathcal{E}^\mathbb{E}}\sum_{j \in \mathcal{R}_{u_e} \setminus \{i_e\}} -log(\sigma(x_{u_e,j}-x_{u_e,i_e})).
\end{align}
Although this recommendation editing loss is quite simple, we find that it meets the basic requirements of recommendation editing and can be widely used in different types of recommendation algorithms. Next, we will demonstrate its effectiveness through experiments.


\section{Experiments}\label{sect:exp}
In this section, we conduct comprehensive experiments to answer the following research questions:
\begin{itemize}[leftmargin=.8cm]
    \item[\textbf{RQ1:}] How effective are the various editing methods in terms of three editing goals?
    \item[\textbf{RQ2:}] How quickly can editing methods rectify recommendation errors?
    \item[\textbf{RQ3:}] What is the impact of the choice of editing pairs on the performance of editing methods?
    \item[\textbf{RQ4:}] How should an appropriate objective for editing methods be determined?
    
\end{itemize}

\subsection{Experimental Setup}
\subsubsection{Datasets}
In our study, we extensively utilize three datasets frequently employed in research on recommendation systems, each distinguished by varying types of negative feedback. This approach is adopted because the intensity of negative feedback in real-world applications tends to be unpredictable and must be tailored according to specific business needs.

\textbf{Epinions}\footnote{https://cseweb.ucsd.edu/~jmcauley/datasets.html}~\cite{zhao2014leveraging}, sourced from online consumer review platform \url{Epinions.com}, comprises user ratings with a range from 1 to 5, detailed item reviews, and directed trust relationships among users. We classify interactions as positive feedback if ratings exceed 3, and negative feedback if ratings are 3 or lower. 

\textbf{KuaiRand}\footnote{https://kuairand.com/}~\cite{gao2022kuairand}, an unbiased sequential recommendation dataset, is derived from the recommendation logs of Kuaishou, a popular video-sharing mobile application. A user clicking on a video indicates positive feedback, while negative feedback is represented by items shown to a user who then clicks the dislike button.

\textbf{QB-video}\footnote{https://static.qblv.qq.com/qblv/h5/algo-frontend/tenrec\_dataset.html}~\cite{yuan2022tenrec} is originated from the QQ browser. The positive feedback here is represented by the user clicking on videos, and negative feedback occurs when a user is shown a video but does not click on it.

The statistics of these datasets are shown in Table \ref{tab:stats}. To ensure the quality of the datasets, we follow the widely used 10-core setting~\cite{wang2019neural} for dataset preprocessing. 

\begin{table}[t]
  \centering
  \small
  \caption{Statistics of the experimental datasets.}
  \resizebox{\linewidth}{!}{
    \begin{tabular}{c|c|c|c|c|c}
    \toprule
    Dataset & \# Users & \# Items & \# Positive Feedback & \# Negative Feedback & \# Negative Feedback Type  \\
    \midrule
    Epinions& 17,871 & 17,453 & 301,378 & 112,396 & Low-rated \\
    KuaiRand& 16,175 & 4,144 & 116,679 & 146,421 & Dislike\\
    QB-video & 27,941 & 15,383 & 1,489,259 & 485,824 &  Exposure \& Unclick \\
    \bottomrule
    \end{tabular}%
  }
  \label{tab:stats}%
  \vspace{-5mm}
\end{table}%

\begin{table*}[t]
  \centering

  \caption{Editing effectiveness of all editing methods. The table presents the reported values for the indicators ES, EC, and EP, expressed as percentages. The best results are in bold, and the second-best results are underlined.Different types of methods are highlighted with different colors(\textcolor{green}{Regularization-based}, \textcolor{blue}{Replay-based}, \textcolor{yellow}{Optimization-based}, \textcolor{orange}{CF with Negative Feedback}, \textcolor{purple}{Model Editing} and \textcolor{red}{Finetuning-based}).}
  \resizebox{\linewidth}{!}{
\begin{tabular}{c|cccccc|cccccc|cccccc}
\toprule
Epinions     & \multicolumn{6}{c|}{MF}           & \multicolumn{6}{c|}{LightGCN}     & \multicolumn{6}{c}{XSimGCL}       \\
Method       & ES & EC & EP & Recall & NDCG & EA & ES & EC & EP & Recall & NDCG & EA & ES & EC & EP & Recall & NDCG & EA \\ 
\midrule

\textcolor{green}{LWF} & 72.37 & 65.54 & 80.80 & 0.1308 & 0.0605 & 1.00 & 57.79 & 44.94 & 80.94 & 0.1569 & \underline{0.0743} & 1.00 & 65.35 & 59.49 & 72.50 & 0.1688 & 0.0834 & 1.00\\
\textcolor{green}{L2} & 72.39 & 66.10 & 80.01 & 0.1306 & 0.0605 & 1.00 & \underline{62.66} & 48.95 & \textbf{87.02} & 0.1540 & 0.0734 & 0.98& \underline{65.93} & 50.54 & \textbf{94.81} & \textbf{0.1737} & \textbf{0.0851} & 0.96\\
\textcolor{green}{SRIU} & 72.10 & 65.64 & 79.97 & 0.1306 & 0.0604 & 1.00 & 56.55 & 44.98 & 76.12 & \underline{0.1570} & 0.0741 & 1.00 & 65.32 & 60.68 & 70.73 & 0.1680 & 0.0829 & 1.00\\
\textcolor{blue}{RSR} & \underline{83.56} & 79.36 & 88.23 & 0.1301 & 0.0595 & 1.00 & 55.04 & 58.00 & 52.36 & 0.1367 & 0.0653 & 1.00 & 65.27 & 58.53 & 73.76 & 0.1691 & 0.0835 & 1.00\\
\textcolor{blue}{SPMF} & \textbf{84.21} & \underline{79.76} & \underline{89.19} & 0.1300 & 0.0596 & 1.00 & 55.45 & 57.68 & 53.38 & 0.1373 & 0.0657 & 1.00 & 65.90 & 59.32 & 74.11 & 0.1696 & 0.0836 & 1.00\\
\textcolor{yellow}{SML} & 0.20 & \textbf{82.30} & 0.10 & 0.0021 & 0.0008 & 1.00 & 20.68 & \underline{69.07} & 12.16 & 0.0776 & 0.0355 & 0.70& 18.04 & \textbf{69.83} & 10.36 & 0.0779 & 0.0356 & 0.80\\
\textcolor{orange}{SiReN} & 57.77 & 53.59 & 62.65 & 0.1228 & 0.0566 & 1.00 & 39.66 & 61.35 & 29.30 & 0.0945 & 0.0445 & 1.00 & 51.48 & 54.06 & 49.14 & 0.1479 & 0.0715 & 1.00\\
\textcolor{orange}{SiGRec} & 0.14 & 0.07 & \textbf{93.42} & 0.1304 & 0.0602 & 1.00 & 39.18 & 66.82 & 27.72 & 0.1036 & 0.0490 & 0.96& 55.91 & \underline{67.54} & 47.70 & 0.1484 & 0.0738 & 0.99\\
\textcolor{purple}{EGNN} & 54.56 & 56.29 & 52.93 & 0.1181 & 0.0548 & 1.00 & 54.46 & 58.66 & 50.82 & 0.1389 & 0.0662 & 1.00 & 54.08 & 54.55 & 53.61 & 0.1556 & 0.0761 & 0.99\\
\textcolor{purple}{BiEGNN} & 72.42 & 64.95 & 81.82 & \underline{0.1311} & \textbf{0.0607} & 0.98& 58.15 & 44.70 & \underline{83.18} & \textbf{0.1573} & \textbf{0.0745} & 0.96& 65.17 & 59.68 & 71.78 & 0.1685 & 0.0832 & 1.00\\
\textcolor{red}{FT} & 72.52 & 64.85 & 82.24 & \textbf{0.1312} & \textbf{0.0607} & 0.97& 57.93 & 45.80 & 78.79 & 0.1564 & 0.0741 & 1.00 & 65.36 & 59.32 & 72.76 & 0.1692 & 0.0835 & 1.00\\
\textcolor{red}{EFT} & 72.50 & 65.02 & 81.93 & 0.1310 & 0.0606 & 0.98& \textbf{75.46} & \textbf{69.20} & 82.96 & 0.1563 & 0.0741 & 1.00 & \textbf{70.35} & 58.13 & \underline{89.06} & \underline{0.1726} & \underline{0.0847} & 1.00\\ \midrule
KuaiRand     & \multicolumn{6}{c|}{MF}           & \multicolumn{6}{c|}{LightGCN}     & \multicolumn{6}{c}{XSimGCL}       \\
Method       & ES & EC & EP & Recall & NDCG & EA & ES & EC & EP & Recall & NDCG & EA & ES & EC & EP & Recall & NDCG & EA \\
\midrule

\textcolor{green}{LWF} & 61.35 & 53.50 & 71.89 & 0.1766 & 0.0657 & 1.00 & 56.51 & 44.28 & 78.06 & \underline{0.2328} & \underline{0.0877} & 1.00 & 56.53 & 42.87 & 82.96 & \underline{0.2493} & \textbf{0.0956} & 1.00\\
\textcolor{green}{L2} & 61.34 & 53.56 & 71.77 & 0.1759 & 0.0654 & 1.00 & \underline{57.09} & 41.29 & \textbf{92.48} & 0.2325 & \textbf{0.0880} & 0.99& 54.88 & 44.90 & 70.57 & 0.2458 & 0.0941 & 1.00\\
\textcolor{green}{SRIU} & 61.29 & 53.50 & 71.74 & 0.1764 & 0.0656 & 1.00 & 55.38 & 44.01 & 74.67 & 0.2326 & 0.0875 & 1.00 & 55.83 & 43.14 & 79.08 & 0.2492 & 0.0952 & 1.00\\
\textcolor{blue}{RSR} & \underline{74.37} & \underline{64.93} & 87.02 & \textbf{0.1808} & \textbf{0.0670} & 1.00 & 49.77 & 48.31 & 51.32 & 0.2033 & 0.0768 & 1.00 & \underline{57.08} & 43.67 & 82.37 & 0.2486 & 0.0954 & 0.96\\
\textcolor{blue}{SPMF} & \textbf{76.76} & \textbf{67.86} & \underline{88.36} & \underline{0.1804} & \underline{0.0668} & 1.00 & 48.90 & 49.15 & 48.65 & 0.1996 & 0.0754 & 1.00 & 57.07 & 43.61 & 82.53 & 0.2489 & \textbf{0.0956} & 0.95\\
\textcolor{yellow}{SML} & 10.38 & 64.86 & 5.64 & 0.0591 & 0.0189 & 0.90& 28.89 & \textbf{63.40} & 18.71 & 0.1649 & 0.0613 & 0.68& 27.44 & \underline{62.79} & 17.56 & 0.1646 & 0.0611 & 0.69\\
\textcolor{orange}{SiReN} & 50.77 & 48.34 & 53.46 & 0.1588 & 0.0587 & 1.00 & 44.08 & 57.72 & 35.65 & 0.1700 & 0.0625 & 1.00 & 37.67 & \textbf{65.62} & 26.42 & 0.1532 & 0.0574 & 1.00\\
\textcolor{orange}{SiGRec} & 0.12 & 0.06 & \textbf{92.47} & 0.1793 & 0.0660 & 1.00 & 33.94 & \underline{61.14} & 23.49 & 0.1536 & 0.0563 & 1.00 & 50.73 & 53.75 & 48.04 & 0.2116 & 0.0815 & 1.00\\
\textcolor{purple}{EGNN} & 52.91 & 46.82 & 60.83 & 0.1695 & 0.0631 & 1.00 & 47.81 & 52.36 & 43.98 & 0.1947 & 0.0736 & 1.00 & 51.19 & 51.19 & 51.19 & 0.2234 & 0.0857 & 0.96\\
\textcolor{purple}{BiEGNN} & 61.28 & 53.63 & 71.48 & 0.1762 & 0.0657 & 1.00 & 56.79 & 44.12 & 79.66 & 0.2326 & 0.0876 & 0.95& 56.05 & 42.72 & 81.48 & 0.2481 & 0.0952 & 1.00\\
\textcolor{red}{FT} & 61.32 & 53.55 & 71.72 & 0.1764 & 0.0656 & 1.00 & 56.65 & 44.21 & 78.82 & \textbf{0.2329} & 0.0875 & 0.96& 56.35 & 42.65 & \underline{83.00} & \textbf{0.2494} & \textbf{0.0956} & 1.00\\
\textcolor{red}{EFT} & 61.38 & 53.58 & 71.83 & 0.1768 & 0.0657 & 1.00 & \textbf{68.90} & 58.97 & \underline{82.86} & 0.2315 & 0.0876 & 1.00 & \textbf{64.33} & 51.92 & \textbf{84.53} & 0.2473 & 0.0951 & 1.00\\ \midrule
QB-video     & \multicolumn{6}{c|}{MF}           & \multicolumn{6}{c|}{LightGCN}     & \multicolumn{6}{c}{XSimGCL}       \\
Method       & ES & EC & EP & Recall & NDCG & EA & ES & EC & EP & Recall & NDCG & EA & ES & EC & EP & Recall & NDCG & EA \\
\midrule

\textcolor{green}{LWF} & 75.21 & 73.90 & 76.57 & 0.2888 & 0.1747 & 1.00 & 60.51 & 47.73 & 82.63 & 0.3185 & 0.1968 & 1.00 & 61.10 & 49.10 & 80.85 & 0.3477 & 0.2159 & 1.00\\
\textcolor{green}{L2} & 76.39 & 70.70 & 83.07 & \textbf{0.2915} & \textbf{0.1761} & 1.00 & \underline{61.74} & 45.35 & \textbf{96.69} & \textbf{0.3193} & \textbf{0.1972} & 0.96& \underline{61.29} & 48.08 & \underline{84.51} & \underline{0.3487} & \underline{0.2169} & 0.91\\
\textcolor{green}{SRIU} & 75.11 & 74.06 & 76.20 & 0.2887 & 0.1744 & 1.00 & 59.19 & 46.51 & 81.39 & 0.3185 & \underline{0.1970} & 1.00 & 60.75 & 54.48 & 68.65 & 0.3383 & 0.2100 & 0.97\\
\textcolor{blue}{RSR} & \underline{80.17} & 73.67 & \underline{87.94} & \underline{0.2910} & 0.1731 & 1.00 & 51.46 & 57.50 & 46.57 & 0.2688 & 0.1605 & 1.00 & 60.01 & 52.07 & 70.81 & 0.3397 & 0.2109 & 0.98\\
\textcolor{blue}{SPMF} & \textbf{80.73} & \underline{80.50} & 80.97 & 0.2855 & 0.1660 & 1.00 & 51.97 & 57.48 & 47.42 & 0.2691 & 0.1611 & 1.00 & 59.97 & 51.76 & 71.27 & 0.3404 & 0.2114 & 0.98\\
\textcolor{yellow}{SML} & 0.02 & \textbf{90.00} & 0.01 & 0.0003 & 0.0002 & 1.00 & 22.98 & \textbf{69.62} & 13.76 & 0.1686 & 0.1015 & 0.85& 19.04 & \textbf{68.70} & 11.05 & 0.1648 & 0.1000 & 0.90\\
\textcolor{orange}{SiReN} & 48.01 & 56.17 & 41.92 & 0.2258 & 0.1317 & 1.00 & 44.40 & 65.31 & 33.63 & 0.1926 & 0.1166 & 1.00 & 42.99 & \underline{65.71} & 31.94 & 0.2100 & 0.1290 & 1.00\\
\textcolor{orange}{SiGRec} & 0.20 & 0.10 & \textbf{90.07} & 0.2882 & 0.1727 & 1.00 & 41.98 & \underline{69.54} & 30.06 & 0.2377 & 0.1398 & 1.00 & 56.20 & 58.94 & 53.71 & 0.3209 & 0.1981 & 1.00\\
\textcolor{purple}{EGNN} & 57.19 & 52.43 & 62.91 & 0.2767 & 0.1652 & 0.95& 53.37 & 57.22 & 50.01 & 0.2814 & 0.1713 & 1.00 & 54.22 & 56.28 & 52.31 & 0.3062 & 0.1890 & 1.00\\
\textcolor{purple}{BiEGNN} & 75.47 & 74.09 & 76.91 & 0.2893 & \underline{0.1751} & 1.00 & 60.22 & 48.49 & 79.42 & 0.3172 & 0.1957 & 1.00 & 60.83 & 49.88 & 77.94 & 0.3456 & 0.2144 & 1.00\\
\textcolor{red}{FT} & 75.69 & 73.70 & 77.79 & 0.2893 & 0.1748 & 1.00 & 60.29 & 48.29 & 80.23 & 0.3182 & 0.1965 & 1.00 & 61.15 & 50.28 & 78.03 & 0.3462 & 0.2148 & 1.00\\
\textcolor{red}{EFT} & 75.29 & 73.94 & 76.70 & 0.2890 & 0.1746 & 1.00 & \textbf{77.95} & 69.22 & \underline{89.20} & \underline{0.3186} & 0.1967 & 1.00 & \textbf{70.71} & 57.33 & \textbf{92.24} & \textbf{0.3502} & \textbf{0.2176} & 1.00\\ \bottomrule
\end{tabular}
}
\label{tab:res}
\vspace{-3mm}
\end{table*}

\subsubsection{Editing Methods}
To evaluate the performance of different methods, we categorize various methods into six categories associated with incremental learning and model editing:

\textbf{Finetuning-based Approaches:} \textbf{FT} (Fine-Tuning)~\cite{Sinitsin2020Editable} and \textbf{EFT} (Editing Fine-Tuning). 
These methods edit the model using a fine-tuning manner based on explicit editing data. For example, the FT method modifies the recommendation model by changing its original parameters, requiring the entire graph's information for graph-based collaborative filtering models. In contrast, EFT modifies the final user and item embeddings through a model that generates these embeddings, without requiring detailed knowledge of the model's architecture and training data.

\textbf{Regularization-based Approaches:} \textbf{LWF} (Learning Without Forgetting)~\cite{li2017learning}, \textbf{L2} (Online L2Reg)~\cite{lin2022continual}, \textbf{SRIU} (Sample Reweighting Incremental Update)~\cite{peng2021preventing}.
These methods also edit models by fine-tuning explicit editing data and using the original model for regularization during the editing process. LWF normalizes the output on the original model using editing data, and L2 utilizes the parameters of the original model for regularization. SRIU contends that overfitting editing data with low confidence in the original model can lead the edited model to deviate from the original model's patterns. It adopts a re-weighting approach that assigns lower weights to data with low confidence.

\textbf{Replay-based Approaches:} \textbf{RSR} (Random Sampling with a Reservoir)~\cite{vitter1985random} and \textbf{SPMF} (Stream-centered
Probabilistic Matrix Factorization model)~\cite{wang2018streaming}. This category method balances the efficiency and effectiveness of model editing by sampling a small number of training examples. The model is edited by fine-tuning based on both the sampled data and the explicit editing data. RSR adopts random sampling, while SPMF uses the original model's prediction scores on training samples to perform weighted sampling.

\textbf{Optimization-based Approaches: SML} (Sequential \\ Meta-Learning) \cite{zhang2020retrain}. SML edits the model by training a Transfer module to generate the new parameters after model editing. Due to the constraints of our task, we cannot train SML on sequential data as originally designed for SML. Instead, we simulate editing data by randomly sampling the recommendation results of the original model multiple times. This simulated data is then used to train the Transfer module for the editing process.

\textbf{CF with Negative Feedback: SiReN} (Sign-aware Recommender system based on GNN models)~\cite{seo2022siren} and \textbf{SiGRec} (Signed Graph Neural Network Recommendation model)~\cite{huang2023negative}. These methods handle both positive and negative feedback during collaborative filtering model training. In this paper, we adopt these approaches to handle negative feedback for model editing purposes.

\textbf{Editing Approach: EGNN} (Editable Graph Neural Networks)~\cite{liu2023editable} and \textbf{BiEGNN} proposed by ourselves. EGNN enhances editing capabilities through an additional MLP (Multi-Layer Perceptron) module. Given the lack of node attributes in collaborative filtering tasks, we use the representations generated by the original model as inputs for the editing module. The existence of two types of nodes (i.e., user and item) utilizing the same editing module may create bias. To tackle this issue, we propose an improved version of EGNN, named BiEGNN, which trains distinct editing modules for users and items respectively to conduct editing.

To verify the generalization of editing methods, we conducted experiments on three representative recommendation models \textbf{MF}(Matrix Factorization)~\cite{rendle2009bpr}, \textbf{LightGCN}~\cite{he2020lightgcn} and \textbf{XSimGCL} (eXtremely Simple Graph Contrastive Learning)~\cite{yu2023xsimgcl} simultaneously.

\subsubsection{Model Editing Setting}
In our experiments, we randomly split the positive feedback in the dataset into a training set and a test set at a ratio of 8:2~\cite{he2020lightgcn}. We use the training set to train the initial collaborative filtering model. 
For the same dataset and backbone, all editing methods underwent testing on the same initial model. 
We used the full set of negative feedback from the dataset as expected editing pairs $\mathcal{E}$, randomly sampling 10 user-item pairs from the intersection of the initial model's recommendation results. The editing pairs consist of the top 50 items predicted for each user and the \textit{correct} editing pairs to form explicit editing pairs $\mathcal{E}^{\mathbb{E}}$. 
Apart from RSR, SPMF, SML, and SiGRec, which perform editing through their unique designs, all other methods employ E-BPR Loss as optimization objectives for editing. To ensure the efficiency of model editing, all finetuning-based methods are restricted to a maximum of 20 rounds of fine-tuning. We evaluated the editing performance of all methods through ES (editing score), EC (editing collaboration), EP (editing prudence), EA (editing accuracy), and Recall and NDCG metrics in the test set after model editing. All methods were repeated 10 times, and the average performance was reported.

\begin{figure}[t]
     \centering
     \includegraphics[width=\linewidth, trim=0cm 0.3cm 0cm 0cm,clip]{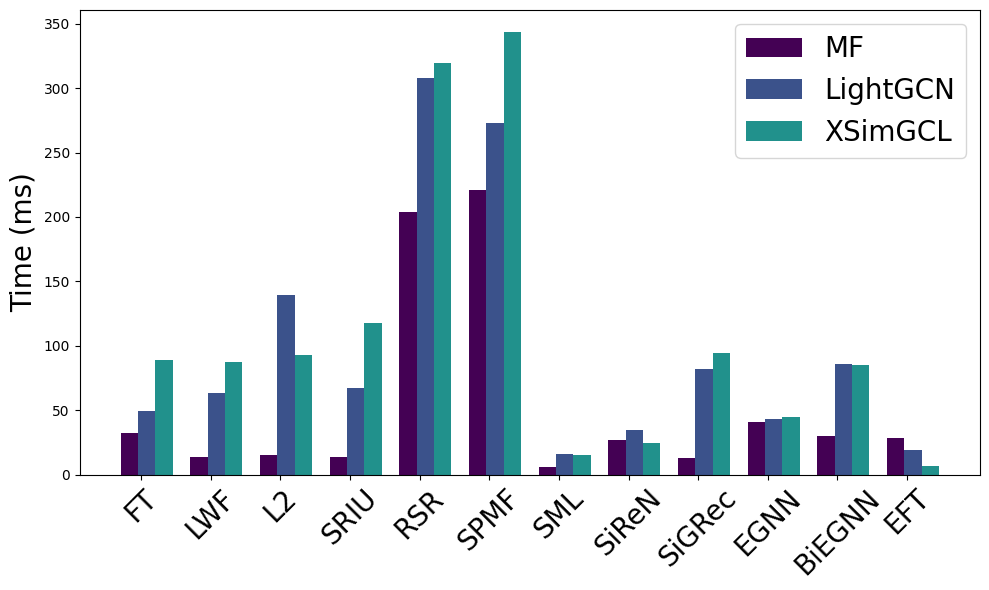}
     \caption{Editing efficiency.}
     \label{fig:time}
     \vspace{-5mm}
\end{figure}
\begin{figure*}
     \centering
     \includegraphics[width=0.9\linewidth, trim=0cm 0cm 0cm 0cm,clip]{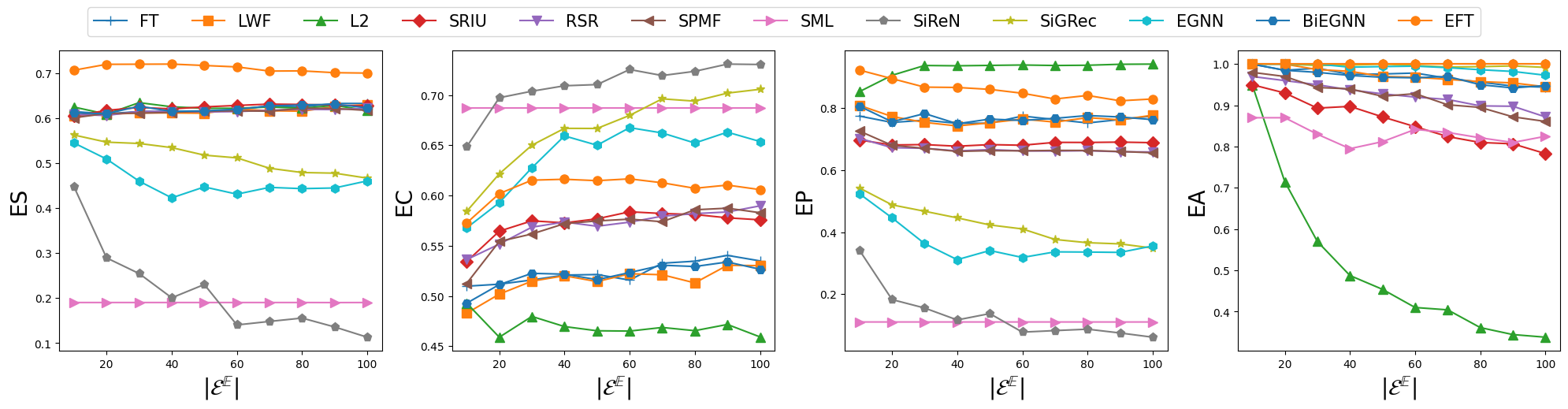}
     \caption{The impact of the number of explicit editing pairs on editing effectiveness.}
     \label{fig:num}
     \vspace{-3mm}
\end{figure*}

\subsection{Editing Effectiveness (RQ1)}\label{sec:edit_effect}
This subsection compares the recommendation editing baselines on the three experimental datasets and three backbone models. The comparison results are reported in Table \ref{tab:res}. We have the following observations:
\begin{itemize}[leftmargin=*]
\item \textbf{Graph is not necessary in model editing}. The EFT method shows superior editing performance in graph-based collaborative filtering methods (LightGCN, XSimGCL), differing from the basic Fine-Tune method only in graph visibility. While graph knowledge usually benefits collaborative filtering, in recommendation editing, the reliance on positive feedback introduces discrepancies with the negative feedback targeted for edits, potentially degrading performance.
\item \textbf{The regularization-based approach improves model editing performance by enhancing Editing Prudence.} The regularization based approach shows suboptimal performance in various recommendation editing tasks. Unlike basic fine-tuning, which lacks constraints, this method uses regularization to minimize divergence from the original model, potentially improving editing caution. However, it can hinder collaborative editing and overly strict constraints may reduce editing accuracy. For example, the L2 method only achieved an editing accuracy of 0.91 on the XSimGCL model using the QB-video dataset, while ideal accuracy approaches 1.00.
\item \textbf{The replay-based approach improves model editing performance by enhancing Editing Collaboration.} Contrary to regularization, the replay-based approach boosts recommendation editing by enhancing collaboration, albeit at the expense of prudence. SPMF and RSR excel in editing MF models but are less effective in graph-based models compared to basic fine-tuning. This approach compensates for missing collaborative information in MF models but magnifies inaccuracies in graph-based models. Notably, SPMF outperforms RSR, suggesting a potential for developing more effective replay strategies for recommendation editing.
\item \textbf{Existing recommendation system methods are not suitable for model editing.} 
We evaluated several recommendation system methods such as SML, SRIU, SiReN, and SiGRec for their compatibility with recommendation editing objectives. Unfortunately, they underperformed, significantly lagging behind basic fine-tuning. This gap likely stems from a mismatch between the methods’ original design goals and the specific demands of recommendation editing.
\item \textbf{Existing model editing methods require targeted design for recommendation editing.} Current model editing methods fall short in recommendation system editing, showing only slight improvements over basic fine-tuning despite targeted enhancements. Developing methods specifically designed for editing recommendation systems is a critical goal in our field.

\item \textbf{The effectiveness of editing is influenced by the intensity of negative feedback.}
From the perspective of different datasets, we can observe that most Editing Methods perform weaker on the KuaiRand dataset compared to the other two. This is because the negative feedback in KuaiRand is characterized by users’ explicit  ``dislike'' actions. Compared to the  ``low-rated'' and  ``exposure \& unclick'' types,  ``dislike'' actions are more intense and targeted, which may result in slightly poorer performance of the editing methods.
\end{itemize}
\subsection{Editing Efficiency (RQ2)}
Recommendation editing efficiency is crucial as it directly impacts our ability to swiftly rectify model issues. This subsection compares the efficiency of various recommendation editing methods across different backbone models. Figure \ref{fig:time} illustrates the editing times for three datasets and models, with the average time noted for each. The EFT method demonstrates superior efficiency in graph-based collaborative filtering models due to its minimal processing of graph information. Conversely, while the regularization-based approach enhances editing speed in MF models, it extends editing times in graph-based models. Similar effects are seen with EGNN and BiEGNN, likely because the tight alignment of user/item representations in these models complicates editing and the additional regularization constraints slow the process. Moreover, the replay-based method’s extra sampling step significantly reduces its efficiency compared to other approaches.

An interesting phenomenon is that time costs for LightGCN+EFT and XsimGCN+EFT are lower than those for MF+EFT. This can be attributed to the insufficient model performance of MF. For LightGCN and XSimGCL, they import explicit collaborative information via graph, rendering the learned embedding distribution more compact. Therefore, the editing method only needs minor modifications to substantially change the target item's order in the candidate set, whereas MF requires more significant modifications.
\subsection{Explicit Editing Pairs Quantity (RQ3)}
In real-world recommendation editing scenarios, the number of explicit editing pairs encountered is often uncertain, and their quantity can significantly impact model editing methods. We explore the impact of the number of explicit editing pairs on the performance metrics of model editing. Figure \ref{fig:num} illustrates how the ES, EC, EP, and EA metrics change for different methods as the number of explicit editing pairs varies, specifically for editing the XSimGCL model trained on the QB-video dataset. From the perspective of ES, the EFT method consistently maintains the best editing capability. Existing well-performing regularization-based and replay-based approaches also remain stable in terms of ES, while poorly performing methods like SML and EGNN see further declines in performance as the number of explicit editing pairs increases. From the perspectives of EC and EP, the majority of methods show a trend of increasing EC and decreasing EP with more explicit editing pairs. This is because more explicit editing pairs can provide more collaborative information but also increase the difficulty of editing, leading to a sacrifice in EP performance. For EA, EFT maintains 100\% accuracy, whereas other methods decline as editing pairs increase. The L2 method uniquely shows opposite trends in EC and EP and suffers the most in EA, likely due to its overly stringent constraints.

\subsection{Editing Objective (RQ4) }
\begin{figure}
     \centering
     \includegraphics[width=\linewidth, trim=0cm 0cm 0cm 0cm,clip]{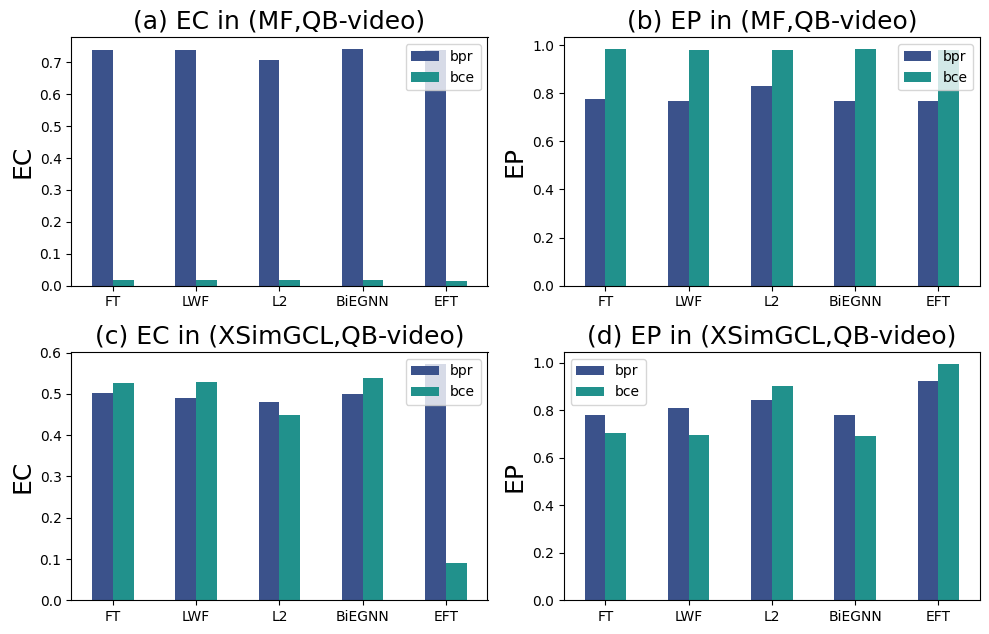}
     \caption{The influence of proper objective.}
     \label{fig:obj}
     \vspace{-4 mm}
\end{figure}

In this subsection, we examined how two common optimization objectives in collaborative filtering, BPR and BCE loss, influence recommendation editing. Figure \ref{fig:obj} depicts the impact on the EC and EP metrics across several prominent editing methods utilizing these objectives. For the MF model, BCE loss offers strong EP performance, yet EC performance is nearly zero. Conversely, BPR loss boosts EC performance significantly by providing positive sample information, albeit with a slight reduction in EP performance by adding collaborative information. In graph-based models, BPR loss yields higher EP performance but slightly worsens EC performance due to the graph’s provision of sometimes inaccurate collaborative information. This effect of BPR loss is amplified by the positive samples, which intensify the inaccuracies from the graph. This observation is corroborated by similar effects seen in the EFT method when using invisible graphs on MF and XSimGCL models.

\section{Conclusion}
In conclusion, this paper introduces the innovative concept of recommendation editing as a critical solution for enhancing user experiences in recommendation systems by efficiently eliminating unsuitable recommendations without necessitating retraining or new data. We established a framework with three core objectives—strict, collaborative, and concentrated rectification—accompanied by three evaluation metrics to thoroughly assess the effectiveness of our approach. The proposed baseline, leveraging editing Bayesian personalized ranking loss, demonstrates significant potential in refining recommendation practices, validated through a comprehensive benchmark of diverse methodologies. Future directions for recommendation editing include: (1) Developing more sophisticated editing algorithms for enhanced editing recommendation performance; (2) Boosting system scalability and computational efficiency; (3) Extending to other recommendation tasks, such as context-aware, cross-domain or sequential recommendation systems, to broaden the applicability; (4) Investigating privacy, fairness, and robustness performance within recommendation editing context.

\clearpage

\bibliographystyle{ACM-Reference-Format}
\bibliography{reference}

\clearpage
\appendix

\section{Experimental Setting}
\subsection{Collaborative filtering model}
The details of the collaborative filtering model we used are as follows:

\textbf{MF} is a cornerstone in collaborative filtering for recommendation systems, decomposing a user-item interaction matrix into lower-dimensional latent factor matrices for users and items. This approach uncovers hidden patterns in data, enabling predictions of unobserved interactions by capturing underlying user preferences and item characteristics. 

\textbf{LightGCN} simplifies the design of Graph Convolutional Networks (GCNs) for collaborative filtering in recommendation systems. It streamlines GCNs by removing feature transformation and nonlinear activation, focusing on the essential component of neighborhood aggregation to learn user and item embeddings. This process effectively captures the high-order connectivity in the user-item interaction graph, enhancing recommendation quality by leveraging the structural information within the graph.

\textbf{XSimGCL} is an improvement upon SimGCL, introducing a highly straightforward graph contrastive learning approach. Not only does it inherit the noise-based augmentation from SimGCL, but it also significantly reduces computational complexity by simplifying the propagation process.

\subsection{Model Parameter Settings}

For MF, LightGCN, and XSimGCL, the embedding size is uniformly set to 64 across all models, and the embedding parameters are initialized using the Xavier method. We optimize the models using Adam with a default learning rate of 0.001 and a default mini-batch size of 2048. The L2 regularization coefficient, $\lambda$, is set to 0.0001. For LightGCN and XSimGCL, the number of model layers is configured to be 3. For XSimGCL, the model parameter lambda is determined through a search within the range [0.01, 0.05, 0.1, 0.2], and eps is searched within [0.05, 0.1, 0.2].

\begin{algorithm}
\caption{EFT with E-BPR}
\begin{algorithmic}[1]
\Require Original model $\mathcal{M}$; explicit editing pairs $\mathcal{E}^{\mathbb{E}}$; set of recommendation result $\mathcal{R}$
\Ensure Edited model $\mathcal{M}^{\mathbb{E}}$
\State $S \leftarrow \emptyset$ 
\For{$(u_e,i_e) \in \mathcal{E}^{\mathbb{E}}$}
    \For{$i \in \mathcal{R}_{u_e}$}
        \If{$i \neq i_e$}
        \State $S \leftarrow S \cup (u_e,i,i_e)$ 
        \EndIf 
    \EndFor
\EndFor
\State Initialize user and item embeddings $P, Q$ from $\mathcal{M}$
\Repeat
    \For{each triplet $(u,i,j) \in S$}
        \State Compute prediction $\hat{x}_{uij} = \hat{x}_{ui} - \hat{x}_{uj}$
        \State Compute BPR loss $\mathcal{L}_{BPR} = -\log(\sigma(\hat{x}_{uij}))$
        \State Update $P_u \leftarrow P_u - \alpha \cdot \frac{\partial \mathcal{L}_{BPR}}{\partial P_u}$
        \State Update $Q_i \leftarrow Q_i - \alpha \cdot \frac{\partial \mathcal{L}_{BPR}}{\partial Q_i}$
        \State Update $Q_j \leftarrow Q_j - \alpha \cdot \frac{\partial \mathcal{L}_{BPR}}{\partial Q_j}$
    \EndFor
    \State Optionally update $\mathcal{M}$ with new $P, Q$
    \State Recompute $\mathcal{R}^{\mathbb{E}}$ using updated $\mathcal{M}$
\Until{$\forall (u_e, i_e) \in \mathcal{E}^{\mathbb{E}}, (u_e, i_e) \notin \mathcal{R}^{\mathbb{E}}$}
\State $\mathcal{M}^{\mathbb{E}} \leftarrow \mathcal{M}$ with updated embeddings $P, Q$
\State \Return $\mathcal{M}^{\mathbb{E}}$
\end{algorithmic}
\label{alg:eft}
\end{algorithm}

\begin{figure*}
     \centering
     \includegraphics[width=\linewidth, trim=0cm 0cm 0cm 0cm,clip]{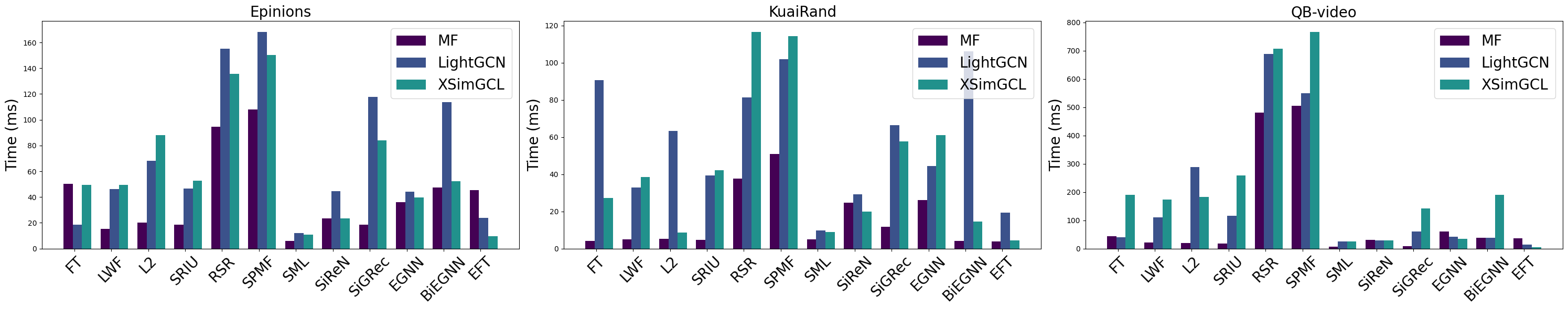}
     \caption{Editing efficiency.}
     \label{fig:time_all}
\end{figure*}

\begin{table*}
  \centering
  \small
  \caption{Experimental results (Mean ± Std.) of all editing methods.}
  \resizebox{\linewidth}{!}{
\begin{tabular}{c|cccc|cccc|cccc}
\toprule
Epinions     & \multicolumn{4}{c|}{MF}           & \multicolumn{4}{c|}{LightGCN}     & \multicolumn{4}{c}{XSimGCL}       \\
Method       & EC & EP & Recall & NDCG  & EC & EP & Recall & NDCG & EC & EP & Recall & NDCG \\ 
\midrule
FT & 0.6485±0.0236 & 0.8224±0.0365 & 0.1312±0.0008 & 0.0607±0.0003 & 0.4580±0.0183 & 0.7879±0.0531 & 0.1564±0.0010 & 0.0741±0.0006 & 0.5932±0.0172 & 0.7276±0.0504 & 0.1692±0.0019 & 0.0835±0.0007\\
LWF & 0.6554±0.0284 & 0.8080±0.0429 & 0.1308±0.0010 & 0.0605±0.0005 & 0.4494±0.0174 & 0.8094±0.0465 & 0.1569±0.0010 & 0.0743±0.0005 & 0.5949±0.0223 & 0.7250±0.0496 & 0.1688±0.0026 & 0.0834±0.0010\\
L2 & 0.6610±0.0247 & 0.8001±0.0354 & 0.1306±0.0006 & 0.0605±0.0003 & 0.4895±0.0117 & 0.8702±0.0183 & 0.1540±0.0006 & 0.0734±0.0004 & 0.5054±0.0213 & 0.9481±0.0141 & 0.1737±0.0004 & 0.0851±0.0001\\
SRIU & 0.6564±0.0268 & 0.7997±0.0457 & 0.1306±0.0012 & 0.0604±0.0006 & 0.4498±0.0134 & 0.7612±0.0556 & 0.1570±0.0013 & 0.0741±0.0006 & 0.6068±0.0208 & 0.7073±0.0511 & 0.1680±0.0019 & 0.0829±0.0010\\
RSR & 0.7936±0.0117 & 0.8823±0.0338 & 0.1301±0.0014 & 0.0595±0.0009 & 0.5800±0.0190 & 0.5236±0.0876 & 0.1367±0.0071 & 0.0653±0.0033 & 0.5853±0.0147 & 0.7376±0.0488 & 0.1691±0.0019 & 0.0835±0.0009\\
SPMF & 0.7976±0.0123 & 0.8919±0.0243 & 0.1300±0.0011 & 0.0596±0.0007 & 0.5768±0.0179 & 0.5338±0.0916 & 0.1373±0.0070 & 0.0657±0.0033 & 0.5932±0.0159 & 0.7411±0.0495 & 0.1696±0.0019 & 0.0836±0.0010\\
SML & 0.8230±0.0001 & 0.0010±0.0000 & 0.0021±0.0000 & 0.0008±0.0000 & 0.6907±0.0001 & 0.1216±0.0000 & 0.0776±0.0000 & 0.0355±0.0000 & 0.6983±0.0000 & 0.1036±0.0000 & 0.0779±0.0000 & 0.0356±0.0000\\
SiReN & 0.5359±0.0331 & 0.6265±0.1316 & 0.1228±0.0068 & 0.0566±0.0036 & 0.6135±0.0771 & 0.2930±0.1859 & 0.0945±0.0377 & 0.0445±0.0182 & 0.5406±0.0491 & 0.4914±0.1653 & 0.1479±0.0203 & 0.0715±0.0100\\
SiGRec & 0.0007±0.0003 & 0.9342±0.0297 & 0.1304±0.0015 & 0.0602±0.0008 & 0.6682±0.0612 & 0.2772±0.1175 & 0.1036±0.0230 & 0.0490±0.0109 & 0.6754±0.0345 & 0.4770±0.1080 & 0.1484±0.0120 & 0.0738±0.0057\\
EGNN & 0.5629±0.0256 & 0.5293±0.0943 & 0.1181±0.0059 & 0.0548±0.0028 & 0.5866±0.0265 & 0.5082±0.0687 & 0.1389±0.0062 & 0.0662±0.0027 & 0.5455±0.0557 & 0.5361±0.1354 & 0.1556±0.0138 & 0.0761±0.0075\\
BiEGNN & 0.6495±0.0275 & 0.8182±0.0387 & 0.1311±0.0008 & 0.0607±0.0004 & 0.4470±0.0239 & 0.8318±0.0194 & 0.1573±0.0008 & 0.0745±0.0003 & 0.5968±0.0199 & 0.7178±0.0593 & 0.1685±0.0028 & 0.0832±0.0013\\
EFT & 0.6502±0.0390 & 0.8193±0.0482 & 0.1310±0.0008 & 0.0606±0.0003 & 0.6920±0.0200 & 0.8296±0.0444 & 0.1563±0.0011 & 0.0741±0.0005 & 0.5813±0.0368 & 0.8906±0.0271 & 0.1726±0.0006 & 0.0847±0.0002\\
 \midrule
KuaiRand     & \multicolumn{4}{c|}{MF}           & \multicolumn{4}{c|}{LightGCN}     & \multicolumn{4}{c}{XSimGCL}       \\
Method       & EC & EP & Recall & NDCG  & EC & EP & Recall & NDCG & EC & EP & Recall & NDCG \\ 
\midrule
FT & 0.5355±0.0333 & 0.7172±0.0131 & 0.1764±0.0015 & 0.0656±0.0006 & 0.4421±0.0135 & 0.7882±0.0201 & 0.2329±0.0013 & 0.0875±0.0006 & 0.4265±0.0166 & 0.8300±0.0331 & 0.2494±0.0018 & 0.0956±0.0006\\
LWF & 0.5350±0.0342 & 0.7189±0.0113 & 0.1766±0.0018 & 0.0657±0.0006 & 0.4428±0.0107 & 0.7806±0.0387 & 0.2328±0.0012 & 0.0877±0.0008 & 0.4287±0.0147 & 0.8296±0.0366 & 0.2493±0.0016 & 0.0956±0.0006\\
L2 & 0.5356±0.0335 & 0.7177±0.0129 & 0.1759±0.0019 & 0.0654±0.0008 & 0.4129±0.0255 & 0.9248±0.0235 & 0.2325±0.0016 & 0.0880±0.0004 & 0.4490±0.0167 & 0.7057±0.0192 & 0.2458±0.0023 & 0.0941±0.0008\\
SRIU & 0.5350±0.0326 & 0.7174±0.0145 & 0.1764±0.0013 & 0.0656±0.0005 & 0.4401±0.0186 & 0.7467±0.0432 & 0.2326±0.0014 & 0.0875±0.0008 & 0.4314±0.0210 & 0.7908±0.0435 & 0.2492±0.0023 & 0.0952±0.0006\\
RSR & 0.6493±0.0362 & 0.8702±0.0327 & 0.1808±0.0013 & 0.0670±0.0006 & 0.4831±0.0169 & 0.5132±0.0592 & 0.2033±0.0079 & 0.0768±0.0029 & 0.4367±0.0184 & 0.8237±0.0370 & 0.2486±0.0020 & 0.0954±0.0006\\
SPMF & 0.6786±0.0377 & 0.8836±0.0323 & 0.1804±0.0012 & 0.0668±0.0004 & 0.4915±0.0161 & 0.4865±0.0656 & 0.1996±0.0084 & 0.0754±0.0033 & 0.4361±0.0187 & 0.8253±0.0382 & 0.2489±0.0018 & 0.0956±0.0007\\
SML & 0.6486±0.0003 & 0.0564±0.0008 & 0.0591±0.0008 & 0.0189±0.0002 & 0.6340±0.0000 & 0.1871±0.0000 & 0.1649±0.0000 & 0.0613±0.0000 & 0.6279±0.0000 & 0.1756±0.0000 & 0.1646±0.0000 & 0.0611±0.0000\\
SiReN & 0.4834±0.0316 & 0.5346±0.1299 & 0.1588±0.0129 & 0.0587±0.0050 & 0.5772±0.0818 & 0.3565±0.1685 & 0.1700±0.0492 & 0.0625±0.0198 & 0.6562±0.0441 & 0.2642±0.1340 & 0.1532±0.0482 & 0.0574±0.0188\\
SiGRec & 0.0006±0.0002 & 0.9247±0.0271 & 0.1793±0.0012 & 0.0660±0.0007 & 0.6114±0.0476 & 0.2349±0.0772 & 0.1536±0.0219 & 0.0563±0.0091 & 0.5375±0.0456 & 0.4804±0.0953 & 0.2116±0.0157 & 0.0815±0.0063\\
EGNN & 0.4682±0.0357 & 0.6083±0.1040 & 0.1695±0.0084 & 0.0631±0.0038 & 0.5236±0.0369 & 0.4398±0.0870 & 0.1947±0.0164 & 0.0736±0.0061 & 0.5119±0.0288 & 0.5119±0.0753 & 0.2234±0.0122 & 0.0857±0.0041\\
BiEGNN & 0.5363±0.0334 & 0.7148±0.0165 & 0.1762±0.0017 & 0.0657±0.0006 & 0.4412±0.0097 & 0.7966±0.0316 & 0.2326±0.0015 & 0.0876±0.0009 & 0.4272±0.0165 & 0.8148±0.0361 & 0.2481±0.0019 & 0.0952±0.0005\\
EFT & 0.5358±0.0338 & 0.7183±0.0142 & 0.1768±0.0016 & 0.0657±0.0005 & 0.5897±0.0212 & 0.8286±0.0275 & 0.2315±0.0013 & 0.0876±0.0004 & 0.5192±0.0237 & 0.8453±0.0109 & 0.2473±0.0004 & 0.0951±0.0002\\
 \midrule
QB-video     & \multicolumn{4}{c|}{MF}           & \multicolumn{4}{c|}{LightGCN}     & \multicolumn{4}{c}{XSimGCL}       \\
Method       & EC & EP & Recall & NDCG  & EC & EP & Recall & NDCG & EC & EP & Recall & NDCG \\ 
\midrule
FT & 0.7370±0.0209 & 0.7779±0.0486 & 0.2893±0.0033 & 0.1748±0.0025 & 0.4829±0.0319 & 0.8023±0.0474 & 0.3182±0.0010 & 0.1965±0.0012 & 0.5028±0.0267 & 0.7803±0.0311 & 0.3462±0.0014 & 0.2148±0.0010\\
LWF & 0.7390±0.0217 & 0.7657±0.0494 & 0.2888±0.0032 & 0.1747±0.0023 & 0.4773±0.0351 & 0.8263±0.0554 & 0.3185±0.0009 & 0.1968±0.0008 & 0.4910±0.0331 & 0.8085±0.0612 & 0.3477±0.0027 & 0.2159±0.0024\\
L2 & 0.7070±0.0131 & 0.8307±0.0282 & 0.2915±0.0019 & 0.1761±0.0015 & 0.4535±0.0187 & 0.9669±0.0070 & 0.3193±0.0002 & 0.1972±0.0002 & 0.4808±0.0270 & 0.8451±0.0696 & 0.3487±0.0019 & 0.2169±0.0014\\
SRIU & 0.7406±0.0255 & 0.7620±0.0506 & 0.2887±0.0028 & 0.1744±0.0020 & 0.4651±0.0264 & 0.8139±0.0357 & 0.3185±0.0007 & 0.1970±0.0006 & 0.5448±0.0359 & 0.6865±0.0428 & 0.3383±0.0031 & 0.2100±0.0024\\
RSR & 0.7367±0.0502 & 0.8794±0.0346 & 0.2910±0.0021 & 0.1731±0.0032 & 0.5750±0.0212 & 0.4657±0.0551 & 0.2688±0.0114 & 0.1605±0.0075 & 0.5207±0.0393 & 0.7081±0.0550 & 0.3397±0.0048 & 0.2109±0.0036\\
SPMF & 0.8050±0.0152 & 0.8097±0.0181 & 0.2855±0.0017 & 0.1660±0.0024 & 0.5748±0.0295 & 0.4742±0.0846 & 0.2691±0.0168 & 0.1611±0.0112 & 0.5176±0.0312 & 0.7127±0.0341 & 0.3404±0.0031 & 0.2114±0.0021\\
SML & 0.9000±0.0000 & 0.0001±0.0000 & 0.0003±0.0000 & 0.0002±0.0000 & 0.6962±0.0000 & 0.1376±0.0000 & 0.1686±0.0000 & 0.1015±0.0000 & 0.6870±0.0000 & 0.1105±0.0000 & 0.1648±0.0000 & 0.1000±0.0000\\
SiReN & 0.5617±0.0449 & 0.4192±0.1517 & 0.2258±0.0464 & 0.1317±0.0311 & 0.6531±0.1108 & 0.3363±0.2369 & 0.1926±0.1093 & 0.1166±0.0678 & 0.6571±0.0849 & 0.3194±0.2502 & 0.2100±0.1086 & 0.1290±0.0690\\
SiGRec & 0.0010±0.0004 & 0.9007±0.0464 & 0.2882±0.0034 & 0.1727±0.0027 & 0.6954±0.0604 & 0.3006±0.0811 & 0.2377±0.0266 & 0.1398±0.0173 & 0.5894±0.0372 & 0.5371±0.0376 & 0.3209±0.0099 & 0.1981±0.0080\\
EGNN & 0.5243±0.0434 & 0.6291±0.0558 & 0.2767±0.0073 & 0.1652±0.0064 & 0.5722±0.0309 & 0.5001±0.0757 & 0.2814±0.0146 & 0.1713±0.0103 & 0.5628±0.0397 & 0.5231±0.0386 & 0.3062±0.0116 & 0.1890±0.0085\\
BiEGNN & 0.7409±0.0231 & 0.7691±0.0300 & 0.2893±0.0017 & 0.1751±0.0011 & 0.4849±0.0224 & 0.7942±0.0289 & 0.3172±0.0021 & 0.1957±0.0021 & 0.4988±0.0383 & 0.7794±0.0843 & 0.3456±0.0038 & 0.2144±0.0028\\
EFT & 0.7394±0.0245 & 0.7670±0.0429 & 0.2890±0.0024 & 0.1746±0.0018 & 0.6922±0.0156 & 0.8920±0.0255 & 0.3186±0.0005 & 0.1967±0.0006 & 0.5733±0.0195 & 0.9224±0.0043 & 0.3502±0.0003 & 0.2176±0.0002\\
 \bottomrule
\end{tabular}
}
\label{tab:dres}
\end{table*}

\subsection{Editing Method}

This section provides a detailed introduction to the editing methods we employed.

\textbf{FT}: 
The FT method transmits the gradient of the edit loss to the original parameters of the collaborative filtering model and then utilizes the Adam optimizer to modify the model parameters for editing.

\textbf{EFT}: 
The EFT method pre-fetches the user and item embeddings generated by the original collaborative filtering model. After parameterizing these embeddings, the gradient of the edit loss is transmitted solely to these embeddings. Editing is then performed by modifying these embeddings. Algorithm \ref{alg:eft} describes the process of editing using the E-BPR Loss with EFT.

\textbf{LWF} and \textbf{L2}: 
LWF and L2 are extensions of the FT method used for editing with E-BPR loss, where an additional regularization loss is introduced to prevent knowledge forgetting. LWF regularizes the editing process by using the MSE loss between the scores of explicit edit pairs from the original and edited models. Conversely, L2 directly employs the MSE loss between the parameters of the original model and the edited model for regularization. The formulations of their edit losses are as follows:
\begin{align}
    \mathcal{L} = \mathcal{L}_{ebpr} + \lambda \mathcal{L}_{reg}.
\end{align}

\textbf{SRIU}: 
SRIU normalizes the output scores of the original model on explicit edit pairs using mean and variance normalization. Then, it computes the weight for each explicit edit pair through Softmax. For the E-BPR loss generated for each triplet (u, $i_{pos}$, $i_{neg}$), the corresponding loss is multiplied by the weight of its associated explicit edit pair.

\textbf{RSR} and \textbf{SPMF}: 
RSR selects $N$ sets of positive feedback at random from historical training data. Conversely, SPMF calculates weights based on the output scores of the original model on the training data and performs weighted sampling to select $N$ sets of positive and negative feedback. Editing is then conducted using the BCE loss.

\textbf{SML}: 
The original SML combines the parameters of the original model with those trained on incremental data as inputs for the parameter transformer. However, in our task, due to the small number of explicit edit pairs, which only contain negative feedback, it is impractical to train a reasonable recommendation model. Therefore, we modify the parameters trained on incremental data to be those obtained from fine-tuning the original model on explicit edit pairs for one round.

\textbf{SiReN} : 
SiReN requires an additional MLP module to process negative feedback. After obtaining the negative feedback embeddings, they are merged with the positive feedback embeddings using an attention mechanism. The MLP module and attention parameters are pre-trained by simulating real negative feedback through random sampling of negative feedback instances.

\textbf{SiGRec}:
SiGRec aggregates negative embeddings through messages passing on a negative graph, which are then concatenated with the user and item embeddings obtained from the original collaborative filtering model. This process results in new user and item embeddings. SiGRec uses the Sign Cosine loss function proposed in the original paper for editing

\textbf{EGNN} and \textbf{BiEGNN}:
EGNN and BiEGNN employ an MLP as the editing module, with the user and item embeddings obtained from the original collaborative filtering model serving as inputs to the editing module. Before editing, it is necessary to pre-train the editing module. This pre-training is conducted by minimizing the score differences between the original model and the modified model on the training data, using the MSE loss.

\subsection{Editing Method Parameter Setting}

All editing methods are optimized using Adam, with the learning rate searched within the range [0.1, 0.03, 0.01, 0.003, 0.001]. For the Regularization-based Approach (LWF, L2), the regularization parameter $\lambda$ is searched within [0.1, 0.01, 0.001, 0.0001]. For the Replay-based Approach (RSR and SPMF), the sample number $N$ is searched within [1, 10, 100].

\subsection{Comparison of different recommendation editing methods.}
Table \ref{tab:cmpedit} show the difference.
We conducted a comparison of various editing methods from four different dimensions, as illustrated in Table \ref{tab:cmpedit}. ``Model Agnostic'' refers to the applicability of the method to models with unknown architectures, ``Training Data Agnostic'' indicates its suitability for models with unknown training data, and ``Extra Storage'' denotes whether additional storage space is required.

\subsection{Hardware and software configuration}
All experiments are executed on a server with 755GB main memory, and two AMD EPYC 7763 CPUs. All experiments are done with a single NVIDIA RTX 4090 (24GB). The software and package version is specified in Table \ref{tab:config}:

\begin{table}[]
    \centering
    \begin{tabular}{c|c}
    \toprule
    Package & Version\\ \midrule
    CUDA & 12.2 \\
    pytorch & 2.1.2\\
    numpy & 1.26.3\\
    scipy & 1.12.0\\
    numba & 0.58.1\\
    \bottomrule
    \end{tabular}
    \caption{Package configurations of our experiments.}
    \label{tab:config}
\end{table}
\vspace{-3mm}

\begin{table}[]
    \centering
    \small
    \begin{tabular}{c|cc|cc|cc}
    \toprule
    Dataset & \multicolumn{2}{c|}{Epinions} & \multicolumn{2}{c|}{KuaiRand} & \multicolumn{2}{c}{QB-video} \\
    Method & Recall & NDCG & Recall & NDCG & Recall & NDCG \\
    \midrule
    MF     & 0.13237 & 0.06145 & 0.18242 & 0.06762 & 0.28626 & 0.16745 \\
    LightGCN & 0.15720 & 0.07468 & 0.23352 & 0.08838 & 0.31929 & 0.19724 \\
    XSimGCL & 0.17399 & 0.08520 & 0.25040 & 0.09628 & 0.35097 & 0.21842 \\
    \bottomrule
    \end{tabular}
    \caption{Recommendation models performance before editing.}
    \label{tab:rec_performences}
    \vspace{-5mm}
\end{table}

\begin{table*}[h]
\caption{Comparison of different recommendation editing methods.}
\begin{tabular}{|l|l|l|l|l|l|}
\hline
Category                                   & Editing Method & Editing Module          & Model Agnostic            & Training Data Agnostic    & Extra Storage             \\ \hline
\multirow{2}{*}{Finetuning-based}          & FT             & Model Parameter         &                           & \checkmark &                           \\ \cline{2-6} 
                                           & EFT            & Model Output            & \checkmark& \checkmark & \checkmark \\ \hline
\multirow{3}{*}{Regularization-based}      & LWF            & Model Parameter         &                           & \checkmark & \checkmark \\ \cline{2-6} 
                                           & L2             & Model Parameter         &                           & \checkmark & \checkmark \\ \cline{2-6} 
                                           & SRIU           & Model Parameter         &                           & \checkmark & \checkmark \\ \hline
\multirow{2}{*}{Replay-based}              & RSR            & Model Parameter         &                           &                           & \checkmark \\ \cline{2-6} 
                                           & SPMF           & Model Parameter         &                           &                           & \checkmark\\ \hline
Optimization-based                         & SML            & Model Parameter         &                           & \checkmark&                           \\ \hline
\multirow{2}{*}{CF with Negative Feedback} & SiReN          & Model Parameter         &                           & \checkmark&                           \\ \cline{2-6} 
                                           & SiGRec         & Model Parameter         &                           & \checkmark&                           \\ \hline
\multirow{2}{*}{Model Editing}             & EGNN           & Additional Architecture & \checkmark& \checkmark&                           \\ \cline{2-6} 
                                           & BiEGNN         & Additional Architecture & \checkmark& \checkmark&                           \\ \hline
\end{tabular}
\label{tab:cmpedit}
\end{table*}

\section{More Experimental Results}
\subsection{Recommendation Models Performance Before Editing}\label{sect:pre-performance}

Table \ref{tab:rec_performences} presents the performance of recommendation models before editing.Although editing methods might cause a slight degradation in the original performance of the recommendation model, severely harmful recommended items are likely to lead to user churn and harm user retention. Therefore, we consider the minor performance loss due to recommendation editing to be an acceptable trade-off.

\subsection{Editing Effectiveness}
Table \ref{tab:dres} displays a detailed comparison of various methods across different datasets and backbones, showcasing the EC, EP, Recall, and NDCG metrics, we report the average value and standard deviation of 10 runs. We can find that the results are consistent with the main analysis in Section~\ref{sec:edit_effect}.

\subsection{Editing Efficiency}

Figure \ref{fig:time_all} illustrates a detailed comparison of editing times for different editing methods across various datasets and backbones. The EFT method maintains exceptionally high editing efficiency across different datasets and backbones.

\section{More discussion on Recommendation Editing}
\subsection{Necessity of Recommendation Editing in practical applications}

In practical industrial recommendation systems, the focus is more on long-term user engagement (such as user retention) rather than just whether a user clicks on more content in the short term\cite{zou2019reinforcement}. The purpose of Recommendation Editing is to improve user experience by correcting inappropriate items in recommendation results, especially discriminatory and illegal ones, thereby avoiding damage to the long-term benefits of the recommendation system. Consider, for instance, if content that severely contradicts a user’s religious beliefs is recommended, it can significantly alienate the user. Upon receiving user feedback, we need to quickly correct our recommendations to prevent such inappropriate content from appearing in future browsing sessions. Otherwise, if a user repeatedly encounters inappropriate content, they are likely to discontinue interacting with the platform and may even choose to stop using it altogether.
\subsection{Application of Recommendation Editing in Industrial Recommendation Systems}
Recommendation Editing can effectively function as a policy and be integrated into the reranking phase of multi-stage recommendation systems, which are commonly used in the industry. The rerank stage determines the item list that the recommendation system ultimately presents to the user. In the rerank stage, multiple policies (such as diversity) are often considered comprehensively to adjust the RankScore generated in the final rank stage\cite{huang2021sliding}:
$$
RS_{rerank}=RS*\prod_{i}(\alpha_i * PolicyScore_i)^{\beta_i},
$$
where $RS_{rerank}$
is the score that determines the final recommendation result, $RS$
 is the score generated in the final rank stage, $PolicyScore_i$
 is the correction score provided by the i-th policy, 
, $\alpha_i$,$\beta_i$
 is the hyperparameters that control the weight of the i-th policy. The recommended editing task can be easily encapsulated into a policy to adjust the final RankingScore. A simple way is to normalize the rankings of each item after editing and use them as PolicyScore:
 $$PolicyScore_{edit}=\frac{candidateNum}{candidateNum + rank_{edit}},$$
 where $candidateNum$
 is the size of candidate items in the rerank stage, and $rank_{edit}$
 is the ranking after editing.
\subsection{Further discussion on limitation of Recommendation Editing in practical applications}
While our work on recommendation editing introduces innovative strategies for enhancing recommendation systems, it is crucial to recognize several limitations and potential negative impacts associated with this approach.

First, the risk of editing attacks poses a significant concern. Malicious actors could exploit the recommendation editing process by introducing false negative signals, tricking the system into misclassifying relevant items as undesirable. This could lead to the unjust removal of valuable content from recommendation lists, thereby degrading the overall quality of recommendations and undermining user trust in the system.

Second, the potential for fairness issues must be considered. The ability to edit recommendations raises the possibility of biased outcomes, where certain items or content types are unfairly favored over others. If not carefully managed, this could result in the marginalization of diverse perspectives and content, leading to a less inclusive recommendation environment. Ensuring that recommendation editing processes are transparent and equitable is essential to mitigate this risk.

Finally, the phenomenon of catastrophic forgetting poses a significant risk. In dynamically evolving environments, continual edits to the recommendation system could inadvertently cause the system to forget previously learned preferences or associations. This could diminish the system's ability to provide consistent and reliable recommendations over time, especially in cases where long-term user preferences are critical.

\end{document}